\documentclass[10pt]{article}
\usepackage{amsfonts}
\usepackage{amsmath}
\usepackage{amssymb}
\usepackage{graphicx}
\usepackage{color}
\usepackage{cite}

\def \be {\begin{equation}}
\def \ee {\end{equation}}
\def \bea {\begin{eqnarray}}
\def \eea {\end{eqnarray}}
\def \nn {\nonumber}
\def \la {\langle}
\def \ra {\rangle}
\def \rr {\raise.35ex\hbox{\small $\prime$}\kern-.17em{\mbox{\large $\imath$}}}
\def \del {\partial}
\def \dels {\partial\kern-.5em / \kern.5em}
\def \As {{A\kern-.5em / \kern.5em}}
\def \Ds {D\kern-.7em / \kern.5em}

\def \a {\alpha}

\def \d {\delta}
\def \eps {\epsilon}

\def \lam {\lambda}
\def \Lam {\Lambda}

\def \mh {\underline{\mu}}
\def \nh {\underline{\nu}}
\def \lh {\underline{\lam}}
\def \kh {\underline{\kappa}}
\def \sh {\underline{\sigma}}
\def \rh {\underline{\rho}}

\newcommand{\cN}{\mathcal{N}}
\newcommand{\cM}{\mathcal{M}}
\newcommand{\vn}{{\vec n}}

\newcommand{\ba}{\begin{eqnarray}}
\newcommand{\ea}{\end{eqnarray}}
\newcommand{\cA}{\mathcal{A}}

\newcommand{\mud}{{\dot\mu}}
\newcommand{\nud}{{\dot\nu}}
\newcommand{\lambdad}{{\dot\lambda}}
\newcommand{\md}{{\dot\mu}}
\newcommand{\nd}{{\dot\nu}}
\newcommand{\ld}{{\dot\lambda}}
\newcommand{\vm}{{\vec m}}
\newcommand{\vl}{{\vec \ell}}
\newcommand{\ve}{{\vec e}}

\begin{document}
\begin{titlepage}

\begin{center}
\hfill UT-08-10
\vskip .5in

\textbf{\LARGE M5 from M2}

\vskip .5in
{\large
Pei-Ming Ho$^\dagger$\footnote{
e-mail address: pmho@phys.ntu.edu.tw}, 
Yutaka Matsuo$^\ddagger$\footnote{
e-mail address:
 matsuo@phys.s.u-tokyo.ac.jp}}\\
\vskip 3mm
{\it\large
$^\dagger$
Department of Physics and Center for Theoretical Sciences, \\
National Taiwan University, Taipei 10617, Taiwan,
R.O.C.}\\
\vskip 3mm
{\it\large
$^\ddagger$
Department of Physics, Faculty of Science, University of Tokyo,\\
Hongo 7-3-1, Bunkyo-ku, Tokyo 113-0033, Japan\\
\noindent{ \smallskip }\\
}
\vspace{60pt}
\end{center}
\begin{abstract}

Recently an action based on Lie 3-algebras 
was proposed to describe M2-branes.
We study the case of infinite dimensional Lie 3-algebras 
based on the Nambu-Poisson structure of
three dimensional manifolds. 
We show that the model contains self-dual 2-form 
gauge fields in 6 dimensions, 
and the result may be 
interpreted as the M5-brane world-volume action.

\end{abstract}

\end{titlepage}
\setcounter{footnote}{0}

\section{Introduction}
\label{sec:BL} 

Recently,
Bagger and Lambert \cite{Bagger:2006sk,Bagger:2007jr,Bagger:2007vi} 
and Gustavsson \cite{Gustavsson:2007vu, Gustavsson:2008dy} proposed 
a supersymmetric Lagrangian for multiple M2-branes  
with a Lie 3-algebra \cite{Filippov} as the internal symmetry
\be\label{BLaction}
{\cal L} = -\frac{1}{2} \la D^{\mu}X^I, D_{\mu} X^I\ra 
+ \frac{i}{2} \la\bar\Psi, \Gamma^{\mu}D_{\mu}\Psi\ra 
+\frac{i}{4} \la\bar\Psi, \Gamma_{IJ} [X^I, X^J, \Psi]\ra 
-V(X) + {\cal L}_{CS},
\ee 
where $D_{\mu}$ is the covariant derivative 
\be
(D_\mu  X^I(x))_a = \partial _{\mu} X^I_a -{f^{cdb}}_a A_{\mu c d}(x) X^I_b, 
\ee
$V(X)$ is the potential term defined by 
\be
V(X) = \frac{1}{12}\la [X^I, X^J, X^K], [X^I, X^J, X^K]\ra, 
\ee
and the Chern-Simons action for the gauge potential is 
\be
{\cal L}_{CS} = \frac{1}{2}\epsilon^{\mu\nu\lam}
\left(f^{abcd}A_{\mu ab}\del_{\nu}A_{\lam cd} 
+ \frac{2}{3} f^{cda}{}_g f^{efgb} A_{\mu ab} A_{\nu cd} A_{\lam ef} \right). 
\ee
The indices $I,J,K$ run in $1,\cdots,8$, 
and they specify the transverse directions
of $M2$-brane; 
$\mu,\nu$ run in $0,1,2$, describing the longitudinal directions.
The indices $a,b,c$ take values in $1,\cdots,\mathcal{D}$
 where $\mathcal{D}$ is the number
of generators of the Lie 3-algebra
specified by a set of structure constants $f^{abc}{}_d$.

The fermionic field $\Psi$ is a Majorana spinor in 10+1 dimensions 
satisfying the chirality condition $\Gamma_{012} \Psi = - \Psi$. 
(The SUSY parameter $\eps$ satisfies 
$\Gamma_{012} \eps = \eps$.)
As a result $\Psi$ has 16 real fermionic components, 
equivalent to 8 bosonic degrees of freedom. 
The bosonic fields include 8 $X^I$'s and 3 $A_{\mu}$'s. 
In 2+1 dimensions, ordinarily a gauge potential has 
one propagating degree of freedom. 
However, here the gauge potential $A$ has no 
canonical kinetic term, 
but only a Chern-Simons term, 
and hence it has no propagating degree of freedom. 

The action has $N=8$ maximal SUSY in $d=3$,
and the SUSY transformations are 
\bea
\d X^I_a &=& i\bar{\eps}\Gamma^I \Psi_a, \\
\d \Psi_a &=& D_{\mu}X^I_a \Gamma^\mu\Gamma^I \eps  
- \frac{1}{6} X^I_b X^J_c X^K_d f^{bcd}{}_a \Gamma^{IJK}\eps, \\
\d \tilde{A}_{\mu}{}^b{}_a &=& 
i\bar{\eps}\Gamma_{\mu}\Gamma_I X^I_c \Psi_d f^{cdb}{}_a. 
\eea
The gauge symmetry for the bosonic fields are written as,
\ba
\delta X^I_a= \Lambda_{cd} {f^{cdb}}_{a} X^I_b\,,\quad
\delta \tilde A_\mu{}^b{}_a=\partial_\mu \tilde\Lambda^b{}_a-
\tilde \Lambda^b{}_c \tilde A_\mu{}^c{}_a +\tilde A_\mu{}^b{}_c
\tilde \Lambda^c{}_a\,.
\ea
For the consistency of these symmetries, we need to require
a generalized Jacobi identity (or the fundamental identity)
to the structure constants,
\ba\label{fi}
\sum_{g} {f_{cde}}^g {f_{abg}}^h=\sum_g\left(
{f_{abc}}^g {f_{gde}}^h+{f_{abd}}^g {f_{cge}}^h+{f_{abe}}^g {f_{cdg}}^h
\right)\,.
\ea
When $\mathcal{D}$ is finite and the metric 
$h^{ab}=\langle T^a, T^b\rangle$  for the basis of generators
$T^a$ ($a=1,\cdots, \mathcal{D}$) is positive definite,
the only known examples are (1) trivial algebra $f^{abc}{}_d=0$,
(2) the so-called $\mathcal{A}_4$ \cite{Kawamura:2003cw},
with $\mathcal{D}=4$ and the structure constant ${f^{abc}}_d=i\epsilon_{abcd}$,
and (3) their direct sums.  Many attempts have been made 
to search for further nontrivial examples of 3-algebras
\cite{FigueroaO'Farrill:2002xg, Bandres:2008vf, Ho:2008bn, Papadopoulos:2008sk}. 
It was conjectured in \cite{Ho:2008bn} that 
there exists no other example of finite dimension 
with a positive definite metric. 
This was finally proved in \cite{Papadopoulos:2008sk,Gauntlett:2008uf} very recently.

When the constraints (1) $\mathcal{D}$ is finite 
and/or (2) $h^{ab}$ is positive definite are replaced by
milder constraints,
there are many varieties of  3-algebras which satisfy 
the fundamental identity (see for example
\cite{Awata:1999dz,Ho:2008bn}).
In particular in \cite{Ho:2008bn}, we commented that for
any manifolds  with Nambu-Poisson structure
\cite{Nambu,qNambu,naryLie2,Jacobian,Vaisman}, one can define
$\mathcal{D}=\infty$ positive-definite Lie 3-algebra.

The BLG theory has gained a lot of attention very recently \cite{recent}. 
In this paper, we examine the BLG theory with $\mathcal{D}=\infty$
Lie 3-algebras based on 3 manifolds $\cN$ with Nambu-Poisson structures.
We will show that the field theory on the membrane world volume $\cM$
can be rewritten as field theory on a 6 dimensional manifold
$\cM\times \cN$ whose bosonic components consist of
the self-dual gauge field on $\cM\times \cN$ and scalar fields which
define the embedding. As this is the field content of
an M5 brane \cite{rM51,rM52}, we interpret it 
as a model of M5-brane constructed out of infinitely many M2-branes.
We note that this problem was considered in Basu and Harvey
\cite{Basu:2004ed} in the context of the generalized Nahm equation.
Our result will provide a new prospective to this problem.

\section{Nambu-Poisson manifold and Lie 3-algebra}

We consider $3$ dimensional manifolds  $\cN$ equipped with
the Nambu-Poisson brackets.  The Nambu-Poisson bracket is
a multi-linear map from
$C(\cN)^{\otimes 3}$ to $C(\cN)$ defined as 
\be
\{ f_1, f_2, f_3 \} = \sum_{\mud,\nud,\lambdad}
P_{\mud \nud \lambdad}(y)
\del_\mud f_1 \del_\nud f_2 \del_\lambdad f_3,
\ee
where $P_{\mud,\nud,\lambdad}$ is an anti-symmetric tensor.
We use the coordinate $y^\mud$ ($\mud=1,2,3$) to parametrize
$\cN$.  The Nambu-Poisson bracket needs to satisfy
the fundamental identity,
\be\label{fi:NP}
\{g, h, \{f_1, f_2, f_3\}\} = 
\{\{g, h, f_1\}, f_2, f_3\} + \{f_1, \{g, h, f_2\}, f_3\}
+\{f_1, f_2, \{g, h, f_3\}\}.
\ee
which gives severe constraints on $P_{\mud,\nud,\lambdad}(y)$
(see for example \cite{Vaisman}).

The simplest possible
Nambu-Poisson bracket is 
the Jacobian determinant for 3 variables $y_\mud (\mud = 1, 2, 3)$
\be
\{ f_1, f_2, f_3 \} = \sum_{\mud,\nud,\lambdad}
\eps_{\mud\nud\lambdad} \del_\mud f_1 \del_\nud f_2 
\del_\lambdad f_3.
\ee
This is the classical Nambu bracket.
In general, it is known that a consistent Nambu bracket
reduces to this Jacobian form locally
by the change of local coordinates \cite{Jacobian}.
So we will use it in the following for simplicity.

Nambu-Poisson bracket may be regarded as the definition 
of Lie 3-algebra in the infinite dimensional space $C(\cN)$.
We write the basis of $C(\cN)$ as $\chi^a(y)$ ($a=1,2,\cdots,\infty$).
We define the Lie 3-algebra structure constant by Nambu-Poisson bracket 
\ba
\left\{ \chi^a, \chi^b, \chi^c\right\}:=\sum_{\mud\nud\lambdad}
\epsilon_{\mud\nud\lambdad}
\partial_\mud \chi^a\partial_\nud \chi^b\partial_\lambdad \chi^c
=\sum_d {f^{abc}}_d \chi^d(y). 
\ea
We write the inner product as integration
\ba
(\chi,\phi)= \int_\cN d^p y \,\mu(y) \chi(y)\phi(y). 
\ea
The measure factor $\mu$ is chosen such that the inner product
is invariant under the Nambu-Poisson bracket, namely
\ba
(\left\{f_1, f_2,\chi\right\}, \phi)+
(\chi, \left\{f_1, f_2,\phi\right\})=0\,.
\ea
The inner product between the basis elements $h^{ab}=(\chi^a,\chi^b)$
is called metric. We choose the dual set of basis 
$\chi_a(Y)$ in $C(\cN)$ such that $(\chi_a, \chi^b)=\delta^a_b$.
We write $(\chi_a,\chi_b)=h_{ab}$ and $\sum_b h^{ab}h_{bc}=\delta^a_c$.
The indices of the structure constant can be
changed by contraction of the metric.   In particular,
$\sum_e h^{ae} {f^{bcd}}_e= f^{bcda}$ defines a totally anti-symmetric
4-tensor.  In order to have finite metric, we need to restrict $\cN$
to a compact manifold.  One may, of course, discuss  noncompact
manifolds by appropriate limits of the compact spaces.

\subsection{Examples}

Since we will analyze only the quadratic terms
of the Lagrangian, the
detail  the 3-algebra
will not be so relevant.
However, since it is of some interest to see the
algebra itself explicitly, we present a few examples
where explicit computation is possible.
It will be useful to proceed to analyze the higher order
terms in the future.

\subsubsection{$T^3$ and $\mathbf{R}^3$}

The simplest example of infinite dimensional
3-algebra is given by $T^3$ with radius $R$.
The basis of functions are parametrized by
$\vn \in \mathbf{Z}^3$ as (if we take $\mu=(2\pi R)^{-3}$) 
\ba
\chi^\vn(\vec y) =\exp(2\pi i \vn \cdot \vec y/R)\,,\quad
\chi_\vn(\vec y)= \exp(-2\pi i \vn \cdot \vec y/R)\,.
\ea
The metric and the structure constants are given by 
\ba
&&h^{\vn_1 \vn_2}=\delta(\vn_1+\vn_2),\\
&&{f^{\vn_1 \vn_2 \vn_3}}_{\vn_4} =(2\pi i/R)^3\vn_1\cdot (\vn_2\times \vn_3)
\delta(\vn_1+\vn_2+\vn_3-\vn_4)\,,\\
&&{f^{\vn_1 \vn_2 \vn_3,{\vn_4}}} =(2\pi i/R)^3\vn_1\cdot (\vn_2\times \vn_3)
\delta(\vn_1+\vn_2+\vn_3+\vn_4)\,.
\ea

If we take $R\rightarrow \infty$, we obtain the Lie 3-algebra
associated with  $\mathbf{R}^3$.
The label for the basis becomes continuous and the metric
becomes the delta function.

\subsubsection{$S^3$}

We introduce four variables $y_1,\cdots, y_4$ and the Nambu-Poisson 
bracket defined by
\ba
P= 
-y_1 \partial_2\wedge \partial_3\wedge \partial_4
+ y_2 \partial_1\wedge \partial_3\wedge \partial_4
-y_3 \partial_1\wedge \partial_2\wedge \partial_4
+ y_4 \partial_1\wedge \partial_2\wedge \partial_3\,.
\ea
If we restrict $C(\cN)$ to the linear functions of $y_i$,
it agrees with $\cA_4$. 
We impose a constraint
$
\phi(y):=y_1^2+y_2^2+y_3^2+y_4^2-1=0
$
in $\mathbf{R}^4$ which defines $S^3$.  
This restriction is compatible with the Nambu-Poisson bracket
in a sense
$
\left\{\phi(y)f_1(y), f_2(y), f_3(y)\right\}|_{\phi(y)=0}=0
$ for any $f_i(y)$.

Square integrable functions on $S^3$ are
given  by combinations of $y^{n_1}_1 y^{n_2}_3 y^{n_3}_3 y^{n_4}_4$. 
By the  constraint, whenever powers of $y_4$ higher than 2 appears, we
can reduce it to zero and one.  Therefore the basis of functions
are given as 
\ba
T_{\vn}=y_1^{n_1}y_2^{n_2}y_3^{n_3}\,,\quad
S_{\vn}=y_1^{n_1}y_2^{n_2}y_3^{n_3} y_4\,,\quad
(n_i\geq 0)\,.
\ea
The 3-algebra becomes 
\ba
\left\{T_\vn, T_{\vm}, T_{\vl}\right\}&=& \vn\cdot
(\vm\times \vl)S_{\vn + \vm+\vl-\vec\rho}\,,\nn\\
\left\{T_\vn, T_{\vm}, S_{\vl}\right\}&=&
\vn\cdot(\vm\times\vl)
\left(T_{\vn+\vm+\vl-\vec \rho}
-\sum_{i=1}^3 T_{\vn+\vm+\vl-\vec \rho+2\ve_i}\right)
 - \sum_{i=1}^3 \ve_i(\vn\times \vm) T_{\vn+\vm+\vl-\vec\rho+2\ve_i}
\,,\nn\\
\left\{T_\vn, S_{\vm}, S_{\vl}\right\}&=&
\vn\cdot(\vm\times\vl)
\left(S_{\vn+\vm+\vl-\vec \rho}
-\sum_{i=1}^3 S_{\vn+\vm+\vl-\vec \rho+2\ve_i}\right)- 
\sum_{i=1}^3 \ve_i(\vn\times (\vm-\vl)) S_{\vn+\vm+\vl-\vec\rho+2\ve_i}\,,\nn
\\
\left\{S_\vn, S_{\vm}, S_{\vl}\right\}&=&\vn\cdot(\vm\times\vl)
\left(T_{\vn+\vm+\vl-\vec \rho}
-2\sum_{i=1}^3 T_{\vn+\vm+\vl-\vec \rho+2\ve_i}+
\sum_{i,j} T_{\vn+\vm+\vl-\vec \rho+2\ve_i+2\ve_j}\right)\nn\\
&& -\sum_i \ve_i(\vn\times \vm +\vm\times \vl + \vl\times \vn)
(T_{\vn+\vm+\vl-\vec \rho+2\ve_i}-T_{\vn+\vm+\vl-\vec \rho+4\ve_i})\nn\\
&& +\sum_{i<j} (\ve_i+\ve_j)(\vn\times \vm +\vm\times \vl + \vl\times \vn)
T_{\vn+\vm+\vl-\vec \rho+2\ve_i+2\ve_j}\,,
\ea
where $\vec\rho=\ve_1+\ve_2+\ve_3$.

\section{M2 to M5}

In this section we will show that 
the BLG model with a Nambu-Poisson structure 
on a 3 dimensional manifold 
contains the low energy degrees of freedom 
on an M5-brane. 
Before going on, 
let us count the number of degrees of freedom 
in the bosonic and fermionic sectors in our model. 
The fermion $\Psi$ is a Majorana spinor in 10+1 dimensions 
with a chirality condition, 
and thus it has 16 real fermionic components, 
equivalent to 8 bosonic degrees of freedom. 
For a 5-brane there are 5 transverse directions 
corresponding to 5 scalars $X^i$. 
For an ordinary 2-form gauge field in 6D, 
there are 6 propagating modes. 
But since we do not have the usual kinetic term for $A$, 
but rather a Chern-Simons term, 
there are only 3 propagating modes. 
The low energy effective theory of an M5-brane 
contains the same number of bosonic and fermionic 
degrees of freedom. 
But a salient feature of the M5-brane is that 
the 2-form gauge field is self-dual. 
Hence our major challenge is to show that 
the gauge field of the BLG model is equivalent to 
a self-dual 2-form gauge field.


A comment on the notation:  we will use  $I.J,K$ 
to label the transverse directions to the membrane worldvolume
$\cM$.  We decompose this eight dimensional space as a direct product
of $\cN$ and remaining 5 dimensional space.  We use $\mud,\nud,\lambdad$
to label $\cN$ and $i,j,k$ to label the transverse directions of the M5-brane.

\subsection{Rewriting fields and covariant derivative}

We expand the fields in BL action in terms of
a basis $\{\chi^a(y) \}$ of $C(\cN)$ as 
\ba
&& X^I(x,y)= \sum_{a} X^I_a(x) \chi^a(y)\label{ex}\,,\\
&& \Psi(x,y) = \sum_a \Psi_a (x) \chi^a(y)\,,\\
&& A_{\mu b}(x, y)=\sum_{a}A_{\mu a b}(x) \chi^a(y)\label{ea}\,,
\ea
which show that the original $3$ dimensional fields on M2-branes 
are promoted to 6 dimensional ones.
We note that the gauge potential $A_{\mu b}$ still contains the index $b$
which labels the basis of $C(\cN)$.

The covariant derivative is  
\be
D_\mu X^I(x,y) =\partial _{\mu} X^I(x,y) 
- \left\{\chi^c,\chi^d,\chi^b \right\} A_{\mu c d}(x) X^I_b 
= \del_{\mu} X^I(x, y) - \left\{ A_{\mu d}(x, y), \chi^d(y), X^I(x, y) \right\}. 
\ee

\subsection{Action}

In the following, we will rewrite the M2 action in terms
of the six dimensional fields defined in (\ref{ex}--\ref{ea}).
For this purpose, we decompose the summation in $I,J,K$ to
the summation in $\mud,\nud,\lambdad$ and $i,j,k$.
Without loss of generality, we will take\footnote{
Of course, this is possible only locally on each coordinate patch
for generic $\cN$.}
\be
P^{\md\nd\ld} = \eps^{\md\nd\ld}
\ee
and 
\be
\la f, g \ra = \int d^3 y \; fg.
\ee

\subsubsection{Potential term}

\bea
L_V &=& -\frac{1}{12}\la [X^I, X^J, X^K], [X^I, X^J, X^K]\ra \nn \\
&=& -\frac{1}{12}\la \left([X^\mud, X^\nud, X^\lambdad]^2 + 
3[X^\mud, X^\nud, X^i]^2
+ 3[X^\mud, X^i, X^j]^2 + [X^i, X^j, X^k]^2\right)\ra.
\label{LV}
\eea

As in matrix models \cite{dWHN,BFSS,IKKT},
the first trick to reorganize the 6 dimensional fields
is to rewrite 
\be
X^\mud(x, y) = y^\mud + A^\mud(x,y).
\ee
Thus
\bea
\la [X^\mud, X^\nud, X^\lambdad]^2 \ra 
&=& (\epsilon^{\mud\nud\lambdad})^2 + 12(\del^\mud A^\mud) 
+12(\del^\mud A^\mud)^2  
-6(\del^\mud A^\nud \del^\nud A^\mud) + {\cal O}(A^3) \nn \\
&=& 6 + 6 \eps^{\mud\nud\lambdad} \del_\mud A_{\nud\lambdad} 
+ 3 (\eps^{\mud\nud\lambdad}\del_\mud A_{\nud\lambdad})^2 
- \frac{3}{2} \eps^{\nud \lambdad \dot\omega}\eps^{\mud \dot\pi \dot\epsilon} 
\del_\mud A_{\lambdad \dot\omega} \del_\nud A_{\dot \pi \dot\epsilon} \nn \\
&& + {\cal O}(A^3), \label{Vquad}
\eea
where 
\be
A_{\mud\nud} \equiv \eps_{\mud\nud\lambdad} A^\lambdad. 
\ee 

The gauge transformation of the gauge field $A^\mud$ is given by 
\be
\d A^\mud(x, y) = \sum_{\a} [ f_{1\a}(x, y), f_{2\a}(x, y), y^\mud+A^\mud] 
= \epsilon^{\mud\nud\lambdad} \sum_{\a}\del_\nud f_{1\a} \del_\lambdad 
f_{2\a} + {\cal O}(A). 
\ee
The linear part of the transformation of $A_{\mud\nud}$
becomes 
\be
\d A = d\Lam + {\cal O}(A), 
\qquad A \equiv \frac{1}{2}A_{\mud\nud} dy^\mud dy^\nud, 
\qquad \Lam = \sum_{\a} f_{1\a} \, d f_{2\a}, 
\ee
as that of a standard 2-form gauge field. 
Thus at the quadratic level of the action, 
we define the field strength of $A$ as 
\be
F_{\md\nud\ld} = \del_\md A_{\nd\ld} + \del_\nd A_{\ld\nd} + \del_\ld
 A_{\md\nd}. 
\ee 
Since $\md,\nd,\ld = 1,2,3$, we have 
\be 
F_{\md\nd\ld} = \frac{1}{2} \eps_{\md\nd\ld} 
\eps^{\dot \pi \dot\omega \dot\epsilon} \del_{\dot \pi} A_{\dot\omega 
\dot\epsilon}.
\ee

Now we try to rewrite the quadratic part of the potential 
in terms of the field strength $F$. 
The first term in (\ref{Vquad}) is a constant. 
The 2nd term is a total derivative, unless $F$ is a constant.  
The 3rd term is 
\be
3(\eps^{\md\nd\ld} \del_\md A_{\nd\ld})^2 = 2 {F_{\md\nd\ld}}^2.
\ee
After integration by parts, the 4th term becomes 
\be
- \frac{3}{2} (\eps^{\md \nd \ld} \del_\md A_{\nd\ld})^2 
= - {F_{\md\nd\ld}}^2. 
\ee
Thus, altogether, 
we find up to total derivatives 
\footnote{
Keeping all the total derivatives, it is 
\be
\la [X^{\md}, X^{\nd}, X^{\ld}]^2 \ra 
= \int d^3 y \; \left[
\left(1+\frac{1}{6}\eps^{\md\nd\ld} F_{\md\nd\ld}\right)^2 
+ \del_{\md}( A^{\md}\del_{\nd}A^{\nd} - A^{\nd}\del_{\nd}A^{\md} ) 
+ {\cal O}(A^3) \right]. 
\ee
}
\be
\la [X^\md, X^\nd, X^\ld]^2 \ra = \int d^3 y \; 
(F_{\md\ld\nd}^2 + \mbox{constant} + {\mathcal{O}(A^3)}). 
\ee

We also have the 2nd term in (\ref{LV}) given by 
\be
-\frac{1}{4}[X^\md, X^\nd, X^i]^2 = - \frac{1}{2} (\del_\md X^i)^2
+{\cal O}(AX^2). 
\ee
The 3rd and 4th terms in (\ref{LV}) have no quadratic terms.

\subsubsection{Fermion kinetic term}

The kinetic term for the fermion is easy to compute: 
\ba
\frac{i}{4} \la\bar\Psi, \Gamma_{IJ} [X^I, X^J, \Psi]\ra \sim
\frac{i}{4} \la\bar\Psi, \Gamma_{\md\nd} [y^\md, y^\nd, \Psi]\ra 
=
\frac{i}{4} \epsilon^{\md\nd\ld}
\la\bar\Psi, \Gamma_{\md\nd} \partial_\ld \Psi\ra 
=\frac{i}{2}\la\bar\Psi, \Gamma^\ld \partial_\ld \Psi\ra. 
\ea

\subsubsection{Chern-Simons term}

The Chern-Simons term in the BL action is 
\ba
L_{CS}=\frac{1}{2}\epsilon^{\mu\nu\lambda}\left(
f^{abcd}A_{\mu a b}\partial_\nu A_{\lambda c d}+
\frac{2}{3}{f^{cda}}_{g}f^{efgb} A_{\mu a b}A_{\nu c d} A_{\lambda ef}
\right). 
\ea
It can be rewritten as
\ba
L_{CS}&=&\frac{1}{2}\epsilon^{\mu\nu\lambda}\left(
\la \left\{\chi^a,\chi^b,\chi^c\right\} , \chi^d\ra
A_{\mu a b}\partial_\nu A_{\lambda c d}\right.\nn\\
&&~~~\left.
+\frac{2}{3} \la \left\{
\chi^c, \chi^d, \chi^a
\right\}, \chi_g\ra \la\left\{\chi^e, \chi^f,\chi^g\right\}, \chi^b\ra
A_{\mu a b}A_{\nu c d} A_{\lambda ef}
\right)\nn\\
&=& \frac{1}{2}\epsilon^{\mu\nu\lambda}\Big(
\la\left\{ A_{\mu b} ,\chi^b, \partial_\nu A_{\lambda d}\right\}, \chi^d\ra
+\frac{2}{3}\la\left\{
A_{\nu d}, \chi^d, A_{\mu b}
\right\},\chi_g\ra
\la\left\{
A_{\lambda f},\chi^f, \chi^g
\right\},\chi^b\ra\Big). 
\ea

Let us now focus on the quadratic term.
%
Here we need to introduce the second trick.
In the choice of the basis $\chi^a$, we pick the first three
as
\be
\chi^\md(y) = y^\md.
\ee
The rest of the basis $\chi^a$ correspond to higher oscillations modes. 
\footnote{
The zero mode $\chi^0 = 1$ has no contribution 
to the CS action since the Nambu-Poisson bracket vanishes 
whenever it is present. 
}
They are ignored for the purpose of this paper, 
which is to identify the physical degrees of freedom of 
the low energy effective M5-brane theory. 

For these modes, we change the label $a$ to the label $\md$ of the coordinates
of $\cN$.  We pick up the corresponding mode in $A_{\mu a}$
(rewritten as $A_{\mu \md}$)
in the quadratic terms in $L_{CS}$ and find 
\be
L_{CS}^{quad} 
= - \frac{1}{2}
\int d^3 y\; \epsilon^{\mu\nu\lambda}\eps^{\md\nd\ld}
\del_{\mu} A_{\nu \md}(x,y) 
\del_\nd A_{\lam \ld}(x, y) +\cdots . 
\ee 
The dots 
$\cdots$ represent the terms which involve $A_{\mu a}$'s 
for higher oscillation modes.
In the following we will ignore them for simplicity.

\subsubsection{Kinetic term for $X$}

The kinetic terms for $X^I$'s are 
\ba
(D_\mu X^I)^2&=& (D_\mu X^\nd)^2 + (D_\mu X^i)^2\nn\\
&=& \left(\partial_\mu A^\nd -\left\{
A_{\mu \ld}, y^\ld, y^\nd
\right\}\right)^2 + (\partial_\mu X^i)^2+\cdots\nn\\
&=& \left(\frac{1}{2} \epsilon^{\nd \md\ld} \partial_\mu A_{\md\ld} -
\epsilon^{\ld\nd\md} \partial_\md A_{\mu\ld}
\right)^2 + (\partial_\mu X^i)^2+\cdots\nn\\
&=& \frac{1}{2}(F_{\mu\nd\ld})^2+(\partial_\mu X^i)^2+\cdots\,,
\ea
where we defined the field strength of $A$ 
\be
F_{\mu\nd\ld} \equiv \del_{\mu}A_{\nd\ld} - \del_{\nd} A_{\mu\ld} 
+ \del_{\ld} A_{\mu\nd}. 
\ee
Here again, $\cdots$ represents those terms involving $A_{\mu a}$
($a\neq 1,2,3$) and higher order terms which we ignore in the analysis
in the following.

\subsection{Equivalence to M5-brane low energy theory}

Collecting all relevant terms, the quadratic part of the BL Lagrangian is 
\ba
L^{quad} &=& - \frac{1}{2} \left[ (\del_{\mu} X^i)^2 + (\del_\md X^i)^2 
\right]
+\frac{i}{2} \la
\bar\Psi, (\Gamma^\mu\partial_\mu +\Gamma^\md \partial_\md)\Psi
\ra\nn\\
&&~~~~~~~
- \frac{1}{4} {F_{\mu\nd\ld}}^2 
- \frac{1}{12} {F_{\md\nd\ld}}^2 
- \frac{1}{2} \eps^{\mu\nu\lam} \eps^{\md\nd\ld} \del_{\mu} A_{\nu \md} \del_\nd
 A_{\lam \ld}. 
  \label{bosonquad} 
\ea
The first two terms on the right hand side 
are the standard kinetic terms for free fields living 
on a 6 dimensional space. 
They agree with what we expect for an M5-brane. 
We will now focus our attention on the gauge fields. 

The last term in the Lagrangian can be rewritten as 
\be
- \frac{1}{2} \eps^{\mu\nu\lam} \eps^{\md\nd\ld} \del_{\mu} A_{\nu \md} \del_\nd
 A_{\lam \ld} 
 = - \frac{1}{8} \eps^{\mu\nu\lam}\eps^{\md\nd\ld}F_{\mu\md\nd}
F_{\nu\lam\ld}
 -\frac{1}{4}\eps^{\mu\nu\lam}\eps^{\md\nd\ld} 
 \del_{\mu}(A_{\md\nd}\del_{\nu}A_{\lam\ld}). 
\ee
where
\be
F_{\nu\lam\ld} \equiv 
\del_{\nu}A_{\lam\ld} - \del_{\lam}A_{\nu\ld}. 
\ee
This is different from the expression of a field strength 
for an ordinary 2-form gauge field 
because it misses the term $\del_{\ld} A_{\nu\lam}$. 
Hence $A_{\md\nd}$ is a 2-form gauge field, 
and $A_{\mu \md}$ is a 1-form gauge field with an addition index $\md$. 
The gauge transformations are 
\bea
\d A_{\md\nd}(x, y) &=& \del_\md \Lambda_\nd(x, y) - \del_\nd \Lambda_\md(x, y), \\ 
\d A_{\mu \md}(x, y) &=& \del_{\mu} \Lambda_\md(x, y). 
\eea
Compared with a 2-form gauge field on a 6-dimensional spacetime, 
we are missing the components $A_{\mu\nu}$ 
and the gauge transformation corresponding to 
the other 3 components of the gauge parameter $\Lambda_{\mu}$. 
We will see below how they automatically appear 
when we analyze the equations of motion in more detail. 

One can rewrite the gauge field relevant terms in (\ref{bosonquad}) as 
\be
L^{quad}_A = 
- \frac{1}{4} {F_{\mu\md\nd}}
(F_{\mu\md\nd} + \tilde{F}_{\mu\md\nd}) 
- \frac{1}{12} {F_{\md\nd\ld}}^2, 
\label{bosonquad1} 
\ee
where $\tilde{F}$ is the Hodge dual of the field strength defined by 
\be
\tilde{F}_{\mh\nh\lh} = \frac{1}{6}\eps_{\mh\nh\lh\kh\sh\rh} F_{\kh\sh\rh}. 
\ee
Here $\mh, \nh, \lh = 0,1, \cdots, 5$ are the collective indices 
for both $\mu$ and $\md$.  
In particular, 
\be
\tilde{F}_{\mu\nu \md} = - \frac{1}{2} \eps_{\mu\nu\lam}\eps^{\md\nd\ld} 
F_{\lam \nd\ld}. 
\label{FFtilde}
\ee
The minus sign on the right hand side of (\ref{FFtilde}) comes from 
$\eps_{\mu\nu\md\lam\nd\ld} = - \eps_{\mu\nu\lam\md\nd\ld}
=-\eps_{\mu\nu\lam}\eps_{\md\nd\ld}$. 


The equations of motion derived from (\ref{bosonquad1}) are 
\bea
&\del_{\mh} F_{\mh\md\nd} = 0, \label{EOMf} \\
&\del_{\nd}F_{\nd\mu\md} + \del_{\nu}\tilde{F}_{\nu\mu\md} = 0. 
\label{CSeom} 
\eea

Let us now show that this set of equations of motion 
(\ref{EOMf}), (\ref{CSeom}) 
is equivalent to the free field theory of a self-dual 2-form 
gauge field in 6 dimensions. 
First we focus on (\ref{CSeom}). 
Combined with the Bianchi identity 
(recall that Hodge dual exchanges equation of motion 
with Bianchi identity)
\be
\del_{\nu}\tilde{F}_{\nu\mu\md} + \del_{\nd}\tilde{F}_{\nd\mu\md} = 0, 
\ee
eq. (\ref{CSeom}) gives 
\be 
\del_{\nd}(F_{\mu\md\nd} - \tilde{F}_{\mu\md\nd}) = 0. 
\ee
Hence there exists a 1-form field $B$ such that 
\be \label{dF}
F_{\mu\md\nd}-\tilde{F}_{\mu\md\nd} = \eps_{\md\nd\ld}\del_{\ld} B_{\mu}. 
\ee

Next we consider the equation of motion (\ref{EOMf}). 
Using (\ref{dF}), we find that (\ref{EOMf}) implies 
\be \label{ii}
\del_{\ld} F_{\ld\md\nd} = - \del_{\lam} F_{\lam\md\nd} 
= \frac{1}{2} \eps_{\mu\nu\lam} \eps_{\md\nd\ld} \del_{\lam} 
F_{\mu\nu\ld} - \eps_{\md\nd\ld} \del_{\lam} \del_{\ld} B_{\lam} 
= - \eps_{\md\nd\ld} \del_{\ld} \del_{\lam} B_{\lam}.  
\ee
Since $\md, \nd, \ld$ can only take 3 different values, 
we have 
\be
F_{\md\nd\ld} = \frac{1}{6} \eps_{\md\nd\ld} 
\eps_{\dot{\kappa}\dot{\sigma}\dot{\rho}} F_{\dot{\kappa}\dot{\sigma}\dot{\rho}}, 
\ee
and hence it follows from (\ref{ii}) that 
\be
\del_{\ld} (F_{\dot{\kappa}\dot{\sigma}\dot{\rho}} + 
\eps_{\dot{\kappa}\dot{\sigma}\dot{\rho}}
\del_{\lam} B_{\lam}) = 0. 
\ee
This is solved by 
\be
F_{\dot{\kappa}\dot{\sigma}\dot{\rho}} + 
\eps_{\dot{\kappa}\dot{\sigma}\dot{\rho}}
\del_{\lam} B_{\lam} = f(x) \eps_{\dot{\kappa}\dot{\sigma}\dot{\rho}}, 
\ee
where $f(x)$ is independent of $y$. 
We can set $f(x) = 0$ by absorbing it in $B_{\mu}$. 
Then the $B_{\mu}$'s are shifted by functions of $x$, 
but this will not change the defining equation (\ref{dF}) of $B_{\mu}$. 
Thus, by suitably choosing $B_{\mu}$, we have 
\be
F_{\dot{\kappa}\dot{\sigma}\dot{\rho}} + 
\eps_{\dot{\kappa}\dot{\sigma}\dot{\rho}}
\del_{\lam} B_{\lam} = 0. \label{dF2} 
\ee

Eqs. (\ref{dF}) and (\ref{dF2}) allow us to define 
a self-dual 2-form gauge field $B_{\mu\nu}$ as 
\be
B_{\mu\nu} = -\eps_{\mu\nu\lam} B_{\lam}, \qquad 
B_{\mu\md} = A_{\mu\md}, \qquad 
B_{\md\nd} = A_{\md\nd}. 
\ee
The field strength $dB$ is denoted by $H$ 
and its components are 
\bea
H_{\mu\nu\lam} &=& \del_{\mu} B_{\nu\lam} + \del_{\nu} B_{\lam\mu} 
+ \del_{\lam} B_{\mu\nu} = - \eps_{\mu\nu\lam} \del_{\rho}B_{\rho}, \\
H_{\mu\nu\md} &=& \del_{\mu} B_{\nu\md} - \del_{\nu} B_{\mu\md} 
- \eps_{\mu\nu\lam} \del_{\md} B_{\lam} 
= F_{\mu\nu\md} - \eps_{\mu\nu\lam} \del_{\md} B_{\lam}, \\
H_{\mu\md\nd} &=& \del_{\mu} B_{\md\nd} - \del_{\md} B_{\mu\nd} 
+ \del_{\nd} B_{\mu\md} = F_{\mu\md\nd}, \\
H_{\md\nd\ld} &=& \del_{\md} B_{\nd\ld} + \del_{\nd} B_{\ld\md} 
+ \del_{\ld} B_{\md\nd} = F_{\md\nd\ld}. 
\eea

For a self-dual field theory, 
the 3-form field strength satisfies 
\bea
H_{\mu\nu\lam} = \frac{1}{6} \eps_{\mu\nu\lam}\eps^{\md\nd\ld} H_{\md\nd\ld},
\qquad
H_{\md\nd\ld} = \frac{1}{6} \eps_{\mu\nu\lam}\eps^{\md\nd\ld} H_{\mu\nu\lam}, \\
H_{\mu\nu \md} = -\frac{1}{2} \eps_{\mu\nu\lam}\eps^{\md\nd\ld} H_{\lam \nd\ld}, 
\qquad
H_{\mu \md\nd} = \frac{1}{2} \eps_{\mu\nu\lam}\eps^{\md\nd\ld} H_{\nu\lam \ld}. 
\eea
(The two equations on the same line are equivalent.) 
It is straightforward to check that the self-duality conditions are 
guaranteed by (\ref{dF}) and (\ref{dF2}). 
Thus we have proven that (\ref{dF}) and (\ref{dF2}) 
are equivalent to the free field theory of a self-dual 2-form gauge field 
in 6 dimensions. 
The Lagrangian (\ref{bosonquad1}) can thus be understood 
as a Lagrangian for a self-dual 2-form gauge field in 6D,
and it is different from such Lagrangians in the literature 
\cite{rM51, rM52}.

\section{Remarks}

One may wonder the possibility of constructing other 
M$p$-branes (which should not exist) in M theory 
from multiple M2-branes. 
However, 
even if we had considered  
a higher dimensional manifold ${\cal N}$ with Nambu-Poisson structure, 
due to the decomposability \cite{Jacobian} 
of the Nambu-Poisson bracket, 
locally one can always choose 3 coordinates $\{y^1, y^2, y^3\}$ 
in terms of which the bracket is simply 
\be
\{f, g, h \}= \eps^{\md\nd\ld} \del_{\md}f\; \del_{\nd}g\; \del_{\ld}h.
\ee
Hence the rest of the coordinates ($y^a$ for $a > 3$) of ${\cal N}$ 
will not induce derivatives or gauge field components. 
There can never be more than 3 of the $X^I$'s turning into 
covariant derivatives. 
The decomposability of the Nambu-Poisson bracket is 
thus the mathematical basis of why there are no other M$p$-branes 
with $p \neq 5$. 

In order to understand this statement, it may be instructive to consider
a straightforward extension,
\ba
P=\partial_1\wedge \partial_2 \wedge \partial_3
+\partial_4\wedge \partial_5 \wedge \partial_6
\ea
which would give us a theory on M8-brane.  
This does not work however since this bracket does NOT
satisfy the fundamental identity!  One may easily confirm this
by examining 
\ba
&&\left\{y_1 y_4, y_2,\left\{y_3,y_5,y_6\right\}\right\}=0\,,\qquad
\mbox{but}\nn\\
&&\left\{\left\{y_1 y_4, y_2,y_3\right\}, y_5,y_6\right\}+
\left\{y_3,\left\{y_1 y_4, y_2,y_5\right\},y_6\right\}+
\left\{y_3,y_5,\left\{y_1 y_4, y_2,y_6\right\}\right\}
=1\,.
\nn
\ea
The fact that the fundamental identity is so restrictive 
is helpful here
to restrict the branes of M-theory to M2 and M5.
 
The Nambu-Poisson tensor $P_{\md\nd\ld}$ is reminiscent 
of the Poisson tensor $\theta_{\md\nd}$ which appears 
on a D$p$-brane world volume when there is a constant $B$-field background 
\cite{B}. 
In both the weak and strong $B$-field limit, the noncommutative structure 
on the D-brane world volume can be approximated by 
the Poisson structure. 
By analogy, we suspect that the M5-brane action presented above 
corresponds to a weak or strong $C$-field limit 
with $C_{\md\nd\ld}$ turned on. 
For a finite value of $C_{\md\nd\ld}$, 
we expect that the Nambu-Poisson bracket to be 
replaced by a quantum version. 
In \cite{Ho:2007vk} we proposed a quantum Nambu bracket 
by examining open membrane scattering amplitudes 
in the large $C$-field background. 
However, the fundamental identity is not preserved by the quantum bracket. 
If it is the correct formulation, 
one should interpret it as the Nambu bracket 
after gauge fixing $A_{\mu\nd\ld} = 0$. 

There are obviously many things to be clarified in the future.
In this paper we consider only the quadratic part of the Lagrangian
and ignored higher terms and also the components for $A_{\mu a}$
for $a\neq 1,2,3$.  To study the precise role played by them
would be essential to understand the precise relation between M2 and M5.
It will also be possible to study the opposite direction, to understand
M2 from M5 brane action \cite{rM51, rM52}.  Since M5 action is 
non-polynomial DBI type action, we expect to have a similar non-polynomial
action which generalizes (\ref{BLaction}).
Since the higher powers of the generators of Lie 3-algebra will be inevitable,
one needs to understand global structure 
associated with a given Lie 3-algebra.  
This is related to the problem of the {\em quantum}
Nambu-bracket, a notoriously difficult problem but
many attempts \cite{qNambu,Awata:1999dz,Ho:2007vk} have been made.  
We hope that  our study here would provide
a good hint to this problem.

\section*{Acknowledgment}

We appreciate partial financial support from
Japan-Taiwan Joint Research Program
provided by Interchange Association (Japan)
by which this collaboration is made possible.

The authors thank Kazuyuki Furuuchi, Yosuke Imamura
Takeo Inami, 
Hsien-chung Kao, Xue-Yen Lin, Darren Sheng-Yu Shih, 
and Wen-Yu Wen for helpful discussions. 
P.-M. H. is supported in part by
the National Science Council,
and the National Center for Theoretical Sciences, Taiwan, R.O.C.
Y. M. is partially supported by
Grant-in-Aid (\#20540253) from the Japan
Ministry of Education, Culture, Sports,
Science and Technology.


\vskip .8cm
\baselineskip 22pt
\begin{thebibliography}{99}
\itemsep 0pt

\bibitem{Bagger:2006sk}
  J.~Bagger and N.~Lambert,
  ``Modeling multiple M2's,''
  Phys.\ Rev.\  D {\bf 75}, 045020 (2007)
  [arXiv:hep-th/0611108].

\bibitem{Bagger:2007jr}
  J.~Bagger and N.~Lambert,
  ``Gauge Symmetry and Supersymmetry of Multiple M2-Branes,''
  Phys.\ Rev.\  D {\bf 77}, 065008 (2008)
  [arXiv:0711.0955 [hep-th]].
 
\bibitem{Bagger:2007vi}
  J.~Bagger and N.~Lambert,
  ``Comments On Multiple M2-branes,''
  JHEP {\bf 0802}, 105 (2008)
  [arXiv:0712.3738 [hep-th]].

\bibitem{Gustavsson:2007vu}
  A.~Gustavsson,
  ``Algebraic structures on parallel M2-branes,''
  arXiv:0709.1260 [hep-th].

\bibitem{Gustavsson:2008dy}
  A.~Gustavsson,
  ``Selfdual strings and loop space Nahm equations,''
  arXiv:0802.3456 [hep-th].

\bibitem{Filippov}
 V. T. Filippov, "n-Lie algebras," Sib. Mat. Zh.,26, No. 6, 126Ð140 (1985).

\bibitem{Kawamura:2003cw}
  Y.~Kawamura,
  ``Cubic matrix, generalized spin algebra and uncertainty relation,''
  Prog.\ Theor.\ Phys.\  {\bf 110}, 579 (2003)
  [arXiv:hep-th/0304149].

\bibitem{FigueroaO'Farrill:2002xg}
  J.~Figueroa-O'Farrill and G.~Papadopoulos,
  ``Pluecker-type relations for orthogonal planes,''
  arXiv:math/0211170.

\bibitem{Bandres:2008vf}
  M.~A.~Bandres, A.~E.~Lipstein and J.~H.~Schwarz,
  ``N = 8 Superconformal Chern--Simons Theories,''
  arXiv:0803.3242 [hep-th].

\bibitem{Ho:2008bn}
  P.~M.~Ho, R.~C.~Hou and Y.~Matsuo,
  ``Lie 3-Algebra and Multiple M2-branes,''
  arXiv:0804.2110 [hep-th].

\bibitem{Papadopoulos:2008sk}
  G.~Papadopoulos,
  ``M2-branes, 3-Lie Algebras and Plucker relations,''
  arXiv:0804.2662 [hep-th].

\bibitem{Gauntlett:2008uf}
  J.~P.~Gauntlett and J.~B.~Gutowski,
  ``Constraining Maximally Supersymmetric Membrane Actions,''
  arXiv:0804.3078 [hep-th].
  
\bibitem{Awata:1999dz}
  H.~Awata, M.~Li, D.~Minic and T.~Yoneya,
  ``On the quantization of Nambu brackets,''
  JHEP {\bf 0102}, 013 (2001)
  [arXiv:hep-th/9906248].

\bibitem{Nambu}
  Y.~Nambu,
  ``Generalized Hamiltonian dynamics,''
  Phys.\ Rev.\  D {\bf 7}, 2405 (1973).

\bibitem{qNambu}
For example,
  L.~Takhtajan,
  ``On Foundation Of The Generalized Nambu Mechanics (Second Version),''
  Commun.\ Math.\ Phys.\  {\bf 160}, 295 (1994)
  [arXiv:hep-th/9301111];
  G.~Dito, M.~Flato, D.~Sternheimer and L.~Takhtajan,
  ``Deformation quantization and Nambu mechanics,''
  Commun.\ Math.\ Phys.\  {\bf 183}, 1 (1997)
  [arXiv:hep-th/9602016]:
  J.~Hoppe,
  ``On M-Algebras, the Quantisation of Nambu-Mechanics, and Volume Preserving
  Diffeomorphisms,''
  Helv.\ Phys.\ Acta {\bf 70}, 302 (1997)
  [arXiv:hep-th/9602020];
  E.~Bergshoeff, D.~S.~Berman, J.~P.~van der Schaar and P.~Sundell,
  ``A noncommutative M-theory five-brane,''
  Nucl.\ Phys.\  B {\bf 590}, 173 (2000)
  [arXiv:hep-th/0005026];
    S.~Kawamoto and N.~Sasakura;
  ``Open membranes in a constant C-field background and noncommutative
  boundary strings,''
  JHEP {\bf 0007}, 014 (2000)
  [arXiv:hep-th/0005123];
  Y.~Matsuo and Y.~Shibusa,
  ``Volume preserving diffeomorphism and noncommutative branes,''
  JHEP {\bf 0102}, 006 (2001)
  [arXiv:hep-th/0010040];
  Y.~Kawamura,
  ``Cubic matrix, Nambu mechanics and beyond,''
  Prog.\ Theor.\ Phys.\  {\bf 109}, 153 (2003)
  [arXiv:hep-th/0207054];
  B.~Pioline,
  ``Comments on the topological open membrane,''
  Phys.\ Rev.\  D {\bf 66}, 025010 (2002)
  [arXiv:hep-th/0201257];
  A.~De Castro, M.~P.~Garcia del Moral, I.~Martin and A.~Restuccia,
  ``M5-brane as a Nambu-Poisson geometry of a multi D1-brane theory,''
  Phys.\ Lett.\  B {\bf 584}, 171 (2004)
  [arXiv:hep-th/0306094];
 T.~L.~Curtright and C.~K.~Zachos,
  ``Deformation quantization of superintegrable systems and Nambu  mechanics,''
  New J.\ Phys.\  {\bf 4}, 83 (2002)
  [arXiv:hep-th/0205063];
  T.~Curtright and C.~K.~Zachos,
  ``Classical and quantum Nambu mechanics,''
  Phys.\ Rev.\  D {\bf 68}, 085001 (2003)
  [arXiv:hep-th/0212267];\\
A review article is,
   D.~S.~Berman,
  ``M-theory branes and their interactions,''
  arXiv:0710.1707 [hep-th].



\bibitem{naryLie2}
M. Schlesinger, J. D. Stasheff, 
``The Lie algebra structure of tangent cohomology 
and deformation theory,''
J. Pure Appl. Algebra {\bf 38} (1985), 313.
P. Hanlon, M. L. Wachs, 
``On Lie $k$-algebras,'' 
Adv. in Math. {\bf 113} (1995), 206. 
J. A. Azc\'{a}rraga, A. M. Perelomov, J. C. P\'{e}rez Bueno, 
``New generalized Poisson structures,'' 
J. Phys. {\bf A29} (1996), 627. 
``$n$-ary Lie and associative algebras,'' 
Redn. Sem. Math. Univ. Pol. Torino {\bf 53} (1996), 373.
J. A. Azc\'{a}rraga, J. M. Izquierdo, J. C. P\'{e}rez Bueno, 
``On the higher order generalizations of Poisson structures'', 
J. Phys. {\bf A30} (1997), L607.
R. Ib\'{a}\~{n}ez, M. de Le\'{o}n, J. C. Marrero, D. Mart\'{i}n de Diego, 
``Dynamics of generalized Poisson and Nambu-Poisson brackets,''
J. Math. Phys. {\bf 38} (1997), 2332.
P. W. Michor, A. M. Vinogradov, 
``n-ary Lie and associative algebras,''
Rend. Sem. Mat. Torino, {\bf 53} (1996), 373.


\bibitem{Jacobian}
R. Weitzenb\"{o}k, 
``Invariantentheorie,'' P. Noordhoff, Gr\"{o}ningen, 1923.
Ph. Gautheron, 
``Some remarks concerning Nambu mechanics,''
Lett. in Math. Phys. {\bf 37} (1996), 103.
D. Alekseevsky, P. Guha, 
``On Decomposability of Nambu-Poisson Tensor,''
Acta. Math. Univ. Commenianae {\bf 65} (1996), 1.
R. Ib\'{a}nez, M. de Le\'{o}n, J. C. Marrero, D. M. de Diego, 
``Dynamics of generalized Poisson and Nambu-Poisson brackets,''
J. of Math. Physics {\bf 38} (1997), 2332.
N. Nakanishi, 
``On Nambu-Poisson Manifolds,''
Reviews in Mathematical Physics {\bf 10} (1998), 499.
G. Marmo, G. Vilasi, A. M. Vinogradov, 
``The local structure of n-Poisson and n-Jacobi manifolds,''
J. Geom. Physics {\bf 25} (1998), 141.

\bibitem{Vaisman}
I. Vaisman, 
``A survey on Nambu-Poisson brackets,''
Acta. Math. Univ. Comenianae {\bf 2} (1999), 213.

\bibitem{recent} 
In addition to \cite{Bandres:2008vf,Ho:2008bn,Papadopoulos:2008sk,
Gauntlett:2008uf}, see: 
  S.~Mukhi and C.~Papageorgakis,
  ``M2 to D2,''
  arXiv:0803.3218 [hep-th].
  D.~S.~Berman, L.~C.~Tadrowski and D.~C.~Thompson,
  ``Aspects of Multiple Membranes,''
  arXiv:0803.3611 [hep-th].
  M.~Van Raamsdonk,
  ``Comments on the Bagger-Lambert theory and multiple M2-branes,''
  arXiv:0803.3803 [hep-th].
  A.~Morozov,
  ``On the Problem of Multiple M2 Branes,''
  arXiv:0804.0913 [hep-th].
  N.~Lambert and D.~Tong,
  ``Membranes on an Orbifold,''
  arXiv:0804.1114 [hep-th].
  J.~Distler, S.~Mukhi, C.~Papageorgakis and M.~Van Raamsdonk,
  ``M2-branes on M-folds,''
  arXiv:0804.1256 [hep-th].
  U.~Gran, B.~E.~W.~Nilsson and C.~Petersson,
  ``On relating multiple M2 and D2-branes,''
  arXiv:0804.1784 [hep-th].
  J.~Gomis, A.~J.~Salim and F.~Passerini,
  ``Matrix Theory of Type IIB Plane Wave from Membranes,''
  arXiv:0804.2186 [hep-th].
  E.~A.~Bergshoeff, M.~de Roo and O.~Hohm,
  ``Multiple M2-branes and the Embedding Tensor,''
  arXiv:0804.2201 [hep-th].
  K.~Hosomichi, K.~M.~Lee and S.~Lee,
  ``Mass-Deformed Bagger-Lambert Theory and its BPS Objects,''
  arXiv:0804.2519 [hep-th].
  G. Papadopoulos, 
  ``On the structure of k-Lie algebras,''
  arXiv: 0804.3567 [hep-th].

\bibitem{rM51}
  P.~Pasti, D.~P.~Sorokin and M.~Tonin,
  ``Covariant action for a D = 11 five-brane with the chiral field,''
  Phys.\ Lett.\  B {\bf 398}, 41 (1997)
  [arXiv:hep-th/9701037];
    I.~A.~Bandos, K.~Lechner, A.~Nurmagambetov, P.~Pasti, D.~P.~Sorokin and M.~Tonin,
  ``Covariant action for the super-five-brane of M-theory,''
  Phys.\ Rev.\ Lett.\  {\bf 78}, 4332 (1997)
  [arXiv:hep-th/9701149];

\bibitem{rM52}
    M.~Aganagic, J.~Park, C.~Popescu and J.~H.~Schwarz,
  ``World-volume action of the M-theory five-brane,''
  Nucl.\ Phys.\  B {\bf 496}, 191 (1997)
  [arXiv:hep-th/9701166].
  

\bibitem{Basu:2004ed}
  A.~Basu and J.~A.~Harvey,
  ``The M2-M5 brane system and a generalized Nahm's equation,''
  Nucl.\ Phys.\  B {\bf 713}, 136 (2005)
  [arXiv:hep-th/0412310].

\bibitem{dWHN}
  B.~de Wit, J.~Hoppe and H.~Nicolai,
  ``On the quantum mechanics of supermembranes,''
  Nucl.\ Phys.\  B {\bf 305}, 545 (1988).
\bibitem{BFSS}
  T.~Banks, W.~Fischler, S.~H.~Shenker and L.~Susskind,
  ``M theory as a matrix model: A conjecture,''
  Phys.\ Rev.\  D {\bf 55}, 5112 (1997)
  [arXiv:hep-th/9610043].
\bibitem{IKKT}
  N.~Ishibashi, H.~Kawai, Y.~Kitazawa and A.~Tsuchiya,
  ``A large-N reduced model as superstring,''
  Nucl.\ Phys.\  B {\bf 498}, 467 (1997)
  [arXiv:hep-th/9612115].
  
\bibitem{B}  
  C.~S.~Chu and P.~M.~Ho,
  ``Noncommutative open string and D-brane,''
  Nucl.\ Phys.\  B {\bf 550}, 151 (1999)
  [arXiv:hep-th/9812219].
  C.~S.~Chu and P.~M.~Ho,
  ``Constrained quantization of open string in background B field and
  noncommutative D-brane,''
  Nucl.\ Phys.\  B {\bf 568}, 447 (2000)
  [arXiv:hep-th/9906192].
  V.~Schomerus,
  ``D-branes and deformation quantization,''
  JHEP {\bf 9906}, 030 (1999)
  [arXiv:hep-th/9903205].
  N.~Seiberg and E.~Witten,
  JHEP {\bf 9909}, 032 (1999)
  [arXiv:hep-th/9908142].
  
\bibitem{Ho:2007vk}
  P.~M.~Ho and Y.~Matsuo,
  ``A toy model of open membrane field theory in constant 3-form flux,''
  Gen.\ Rel.\ Grav.\  {\bf 39}, 913 (2007)
  [arXiv:hep-th/0701130].




\end{thebibliography}
\end{document}